\documentclass[aps,twocolumn,prl,reprint,amsmath,amssymb,showpacs,superscriptaddress]{revtex4-1}
\usepackage{multirow}
\usepackage{graphicx,epsfig}
\usepackage{color,soul} 
\usepackage[colorlinks=true,citecolor=blue,linkcolor=red,linktocpage=true,pagebackref=false]{hyperref}
\usepackage[linktocpage=true]{hyperref}

\newcommand{\ba}{\begin{eqnarray}}
\newcommand{\ea}{  \end{eqnarray}}

\def \beq {\begin{equation}}
\def \edq {\end{equation}}
\def \bes {\begin{subequations}}
\def \eds {\end{subequations}}
\def \beqn {\begin{equation*}}
\def \edqn {\end{equation*}}

\begin{document}
\title{Fractional spin and Josephson effect in time-reversal-invariant topological superconductors}
\author{Alberto Camjayi}
\affiliation{Departamento de F\'{\i}sica, FCEyN, Universidad de Buenos Aires and IFIBA, Pabell\'on I, Ciudad Universitaria, 1428 CABA Argentina} 
\author{Liliana Arrachea}
\affiliation{International Center for Advanced Studies, ECyT-UNSAM, Campus Miguelete, 25 de Mayo y Francia, 1650 Buenos Aires, Argentina}
\author{Armando Aligia}
\affiliation{Centro At\'omico Bariloche and Instituto Balseiro, CNEA, 8400 S. C. de Bariloche, Argentina}
\author{Felix von Oppen}
\affiliation{\mbox{Dahlem Center for Complex Quantum Systems and Fachbereich Physik, Freie Universit\"at Berlin, 14195 Berlin, Germany}}

\begin{abstract}
Time reversal invariant topological superconducting  (TRITOPS) wires are known to host a fractional spin  
$\hbar/4$ at their ends. We investigate how this fractional spin affects the Josephson current 
in a TRITOPS-quantum dot-TRITOPS Josephson junction, describing the wire in a model which can be tuned 
between a topological and a nontopological phase. We compute the equilibrium Josephson current of the 
full model by continuous-time Monte Carlo simulations and interpret the results within an effective 
low-energy theory. We show that in the topological phase, 
the $0$-to-$\pi$ transition is quenched via formation of a spin singlet from the quantum dot spin and the 
fractional spins associated with the two adjacent topological superconductors. 
\end{abstract}

\maketitle

{\em Introduction.---}The interplay of many-body interactions in quantum dots and superconductivity has been at the focus of interest for some time \cite{rev1,rev2,exp-jos1,exp-jos2,exp-jos3,pillet}. While electrons are paired in superconductors, the charging energy effectively suppresses pairing in quantum dots. A prominent consequence of this competition is the transition between $0$ and $\pi$ junction behavior of the Josephson current in devices where a quantum dot (QD) connects between ordinary (nontopological) singlet-superconducting wires (S-QD-S junction) \cite{kulik,shiba,matveev,spivak}. As a result of numerous studies \cite{fasepi1,fasepi2,fasepi3,sim,fasepi4,fasepi5,oguri,allub,kirs}, this phenomenon is now well understood for conventional superconductors. Essentially, S-QD-S junctions exhibit $\pi$-junction behavior when the QD hosts an effective spin-$1/2$ degree of freedom. 
 
Here, we address the $0$ to $\pi$ transition for Josephson junctions in which a quantum dot connects between time-reversal-invariant topological superconductors (TRITOPS). Unlike their time-reversal-breaking cousins \cite{maj-rev,kitaev,das-sarma,felix}, TRITOPS preserve time-reversal symmetry and can coexist with an unpolarized quantum-dot spin. It is thus an interesting question whether $\pi$-junction behavior can be observed in TRITOPS-QD-TRITOPS junctions. Such junctions differ from conventional S-QD-S junctions in several ways. First, the Majorana-Kramers pairs present in the topological phase allow for the coherent transfer of single electrons, while the Josephson current of a conventional junction is carried by Cooper pairs. Even more intriguing, TRITOPS host a fractional $\hbar/4$ spin at their ends. Thus, a TRITOPS-QD-TRITOPS junction allows one to study the hybridization of fractional and ordinary spins. We show that the $0$-$\pi$ transition constitutes a signature which distinguishes between the topological and the nontopological phase, and trace the quenching of the transition for TRITOPS to the formation of a spin singlet from the quantum-dot spin and the fractional spins of the adjacent TRITOPS. 

In the wake of proposals to engineer time-reversal-breaking topological phases and corresponding experiments, there has also been substantial interest in time reversal invariant topological superconductors (TRIPTOPS) \cite{tri1,tri2,tri3,fan-kane-mele,kesel,tri4,tri5,tri6,tri8,tri9,tri10,tri11,tri12,tri13}. TRITOPS are characterized by Kramers pairs of Majorana end states and localized fractional spins \cite{kesel}. Time reversal protects the pair of Majoranas from hybridizing which therefore generically remain at zero energy. Similarly, the fractional spin is topologically protected and cannot be determined from a local measurement without breaking time reversal. Several routes have been proposed to engineer TRITOPS although their experimental realization is more demanding than that of time-reversal-breaking topological superconductors \cite{tri12}. 

Conventional Josephson junctions assume their minimal energy at zero phase difference and their maximal energy at a phase difference of $ \pi$ ($0$-junction behavior). This behavior is reversed in 
$\pi$ junctions which assume their minimal energy at a phase difference of $\pi$ \cite{rev1,rev2}. In S-QD-S junctions, $\pi$-junction behavior occurs when the quantum dot forming the junction is singly occupied and acts effectively as a magnetic impurity. When the QD is weakly coupled to the superconductors, tunneling of Cooper pairs between the conventional superconductors relies on a forth-order cotunneling process \cite{rev1,spivak}. This process includes a $\pi$ phase shift which originates from the Fermi statistics of electrons and becomes manifest in the $\pi$-junction behavior. As a consequence, the current-phase relation of the junction phase shifts by $\pi$ when the occupation of the quantum dot is tuned from even to odd. When the quantum dot is strongly coupled to the superconductors, the impurity spin can be screened, turning a doublet into a singlet ground state and resulting in $0$-junction behavior. Depending on the parameter regime, this transition can be described as a result of Kondo correlations or a zero-energy crossing of a Yu-Shiba-Rusinov state \cite{kirs}.
  
{\em Model.---}Our considerations are based on a time-reversal-invariant superconductor with Hamiltonian  \cite{fan-kane-mele}
\ba \label{wires}
  H_{\alpha} & = & \sum_{j=1}^N \sum_\sigma \left(-t c^{\dagger}_{\alpha, j+1, \sigma} c_{\alpha, j,\sigma} + i \lambda_{\sigma} c^{\dagger}_{\alpha, j+1, \sigma} c_{\alpha,j, \sigma}  \right. \nonumber \\
& & \left.  +
\Delta_{\sigma} e^{i \phi_{\alpha}} c_{\alpha,j+1,\sigma}^{\dagger} c^{\dagger}_{\alpha,j, \overline{\sigma}}
+ {\rm H.c.} - \mu\;  n_{\alpha,j, \sigma} \right),
\ea 
where $\lambda_{\uparrow,\downarrow}=\pm \lambda$, $ \Delta_{\uparrow,\downarrow}=\pm \Delta$ and $\overline{\uparrow}=\downarrow, \;\overline{\downarrow}=\uparrow$. Moreover, $t$ is the hopping parameter, $\mu$ the chemical potential, and $\lambda$ and $\Delta$ are the strengths of Rashba spin-orbit coupling and extended s-wave pairing, respectively. The index $\alpha=L,R$ labels the left and right superconductors of the junction with order parameter phases $\phi_\alpha$.  The phase difference $\phi=\phi_{L}-\phi_R=2 \pi \Phi/\Phi_0$ can be tuned by including the junction in a superconducting loop and threading the loop by a magnetic flux $\Phi$. ($\Phi_0= h/2e$ denotes the superconducting flux quantum.) The entire TRITOPS-QD-TRITOPS Josephson junction is then described by the Hamiltonian 
 \beq \label{ham}
 H= \sum_{\alpha=L,R} H_{\alpha} + H_{c}  + H_d.
 \edq
Here, 
\beq
H_d= \varepsilon_d \sum_{\sigma} n_{d, \sigma } + U n_{d \uparrow} n_{d \downarrow}
\edq
describes the quantum dot with gate-tunable level energy $\varepsilon_d$, spin-resolved level occupation $n_{d,\sigma}$, 
and charging energy $U$, and 
\beq
    H_{c} = -t^{\prime}  \sum_{\sigma} \left[ \left(c^{\dagger}_{L, N, \sigma}+c^{\dagger}_{R, 1, \sigma} \right)  
    d_{\sigma} + {\rm H.c.} \right] 
 \edq
 accounts for the hybridization between quantum dot and superconductors. 
 
The Hamiltonian $H_\alpha$ supports topological and nontopological phases. The topological phase occurs when $|\mu|< 2 \lambda$ and is characterized by Kramers pairs of Majorana end states. For each lead, the corresponding Majorana operators can be combined into conventional fermionic operators 
\begin{equation} \label{gammas}
  \Gamma_{L/R} = \int {\rm d}x \varphi_{L/R}(x) [\psi_\uparrow(x)  \mp i \psi^\dagger_\downarrow(x) ],
\end{equation}
where $\varphi_{L/R}(x)$ denotes the Majorana wavefunctions of the left (L) and right (R) superconducting lead and $\psi_\sigma(x)$ denotes the electron field operator for spin $\sigma$
(see Supplementary Material, Sec.\ 2 \cite{suppl}). While the Majorana operators mix the two spin components, the operators $\Gamma_{L/R}$ remove a spin of $\hbar/2$ from one end of  the wire. Thus, $\Gamma_{L/R}$ and $\Gamma_{L/R}^\dagger$ toggle the the system between ground states with fractional spins of $\pm\hbar/4$ localized at the ends of the wire \cite{kesel}.  

  \begin{figure}[t]\begin{center}
  \includegraphics[width=0.8\columnwidth,height=4.5cm]{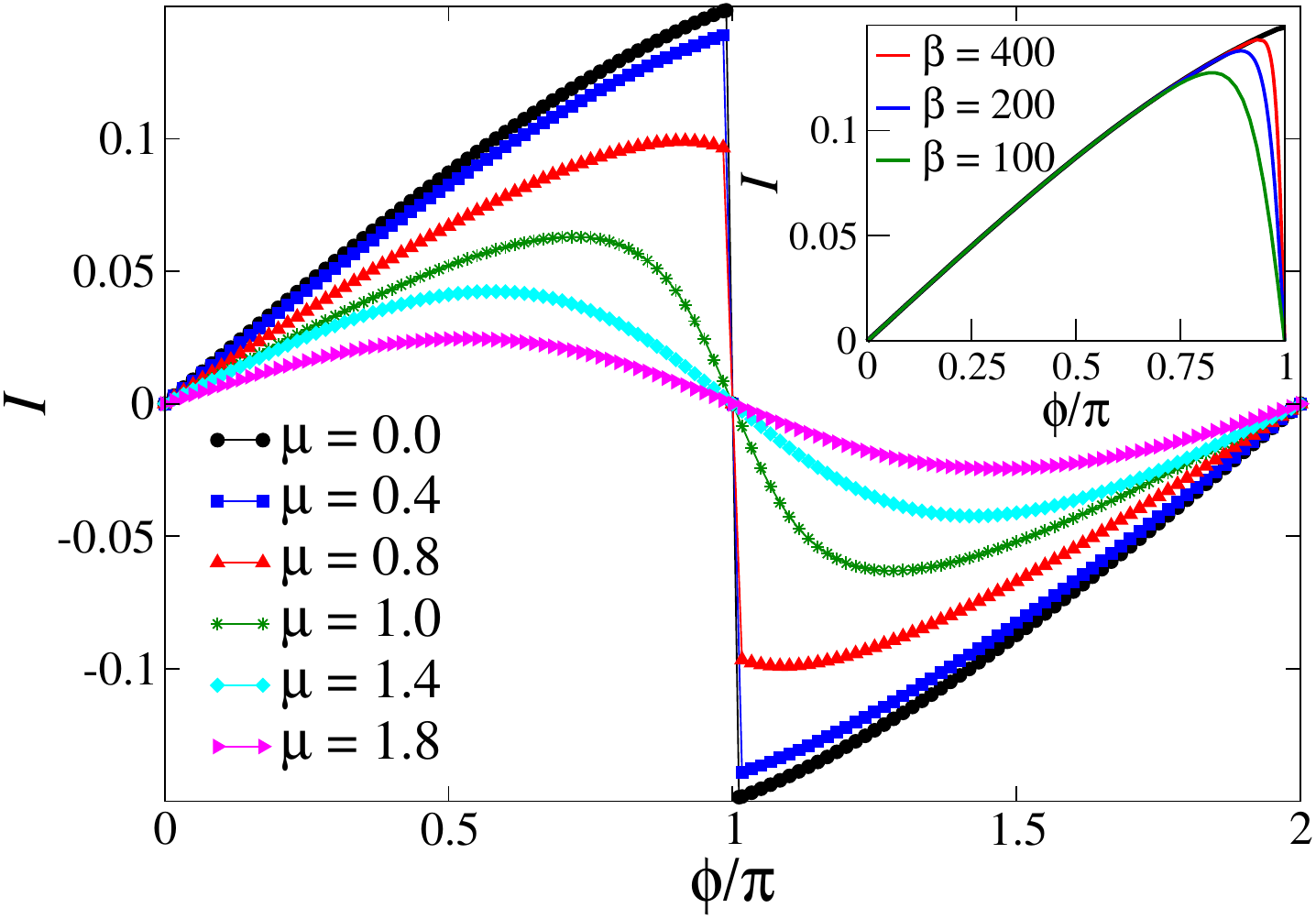}\\
   \includegraphics[width=0.9\columnwidth,height=4.5cm]{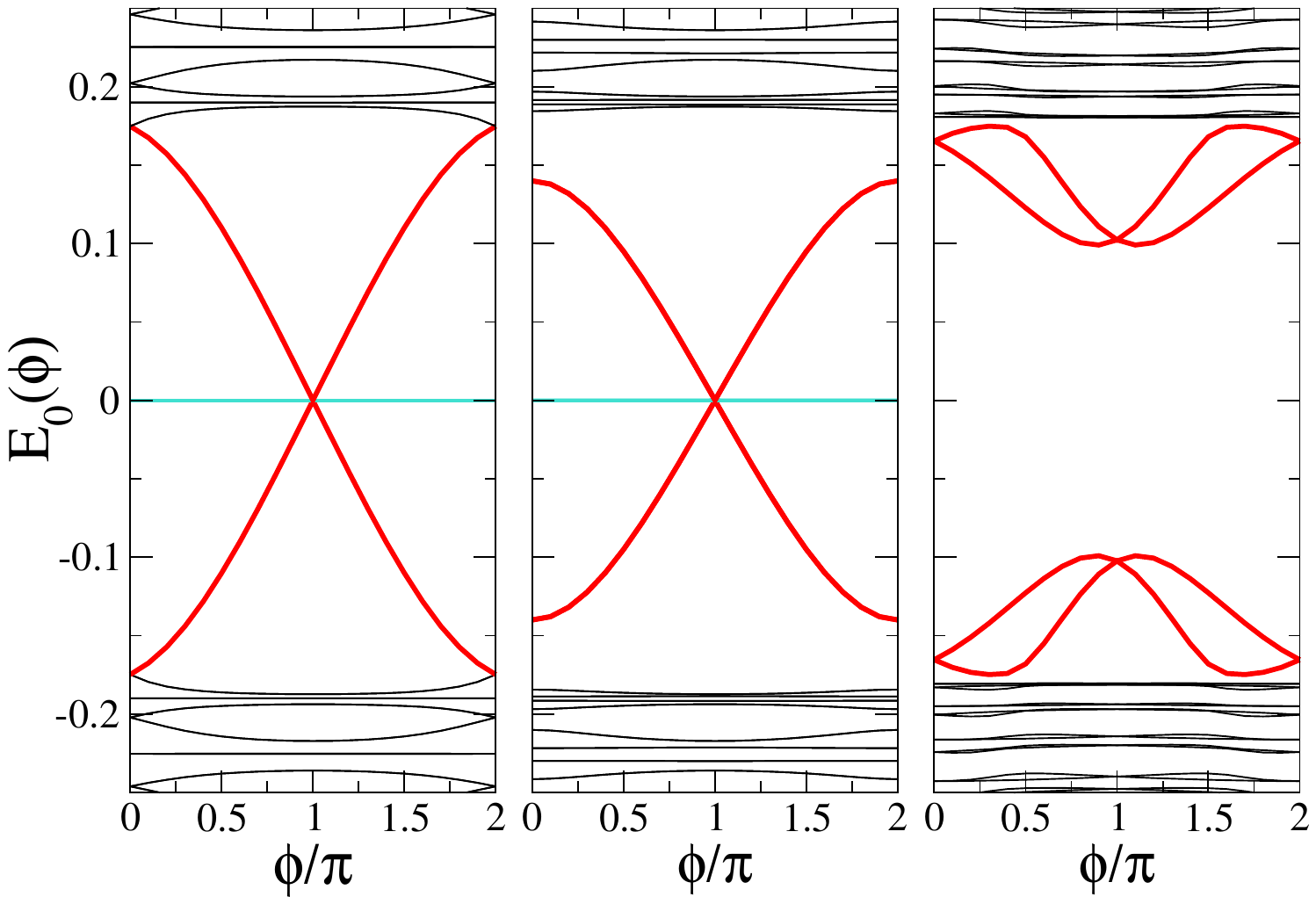}\\
  \caption{(Color online) Top: Josephson current {\em vs} $\phi$ for a quantum dot at $T=0$ with $U=0$, 
  $t^{\prime}= t$, $\lambda=t/2$, $\Delta=t/5$, and values of $\mu$ in the topological ($\mu<t$) as well as the 
  nontopological ($\mu>t$) phase. The wires have $N=500$ sites.
Inset: Josephson current at finite temperature.  
Red, blue, and green lines correspond to $\beta= 400, 200 \mbox{ and } 100$, respectively.
The $T=0$ case is plotted in black for reference.
  Bottom: Spectrum of $\hat{\cal H}_{BdG}$ for $\mu=\varepsilon_d=0$ (left), $\mu=0$, 
  $\varepsilon_d=t$ (middle), and $\mu=-\varepsilon_d=1.8$ (right). 
  Other parameters as in top panel. Energies are measured in units of $t=1$. }\label{plot1}
\end{center}\end{figure}

{\em Numerical results.---}The Josephson current can be computed from the Green function expression
\begin{eqnarray} \label{jcur}
I =\frac{2 t^{\prime 2}}{\beta} \sum_{\sigma} \sum_{n} \mbox{Im}\left[g_{1\alpha,\sigma }^{(12)}(i \omega_n) G^{(21)}_{d,\sigma}( i \omega_n) \right] .
\end{eqnarray}
The derivation is included in Ref.\ \cite{suppl} (see Sec.\ 1). The Green functions correspond to the Matsubara components of frequency $\omega_n =( 2 n +1)\pi /\beta$ ($\beta$ is the inverse temperature) of the imaginary-time Green functions $g_{1\alpha,\sigma }^{(12)}(\tau) = - \langle T_{\tau} \left[ \hat{c}^{\dagger}_{\alpha, 1, \sigma} (\tau)\hat{c}^{\dagger}_{\alpha, 1,  \overline{\sigma}} (0) \right] \rangle_0$ and $G^{(21)}_{d,\sigma}( \tau) =  - \langle T_{\tau} \left[ \hat{d}_{\sigma} (\tau)\hat{d}_{\overline{\sigma}} (0) \right] \rangle$, 
where $\langle \ldots \rangle_0$ ($\langle \ldots \rangle$) denotes the ensemble average over the states of $H_{\alpha}$ ($H$). The first Green function can be obtained exactly. 
 
First consider a junction with a noninteracting quantum dot. For $U=0$, the Green function ${G}_{d, \sigma}(i \omega_n)$ and thus the Josephson current can also be evaluated analytically.  Moreover, our model can be written in Nambu representation with a Bogoliubov de-Gennes (BdG) Hamiltonian $\hat{\cal H}_{BdG}= \hat{ \cal H}_0 \tau_z + \hat{\Delta}  \tau_x$, where $\hat{\cal H}_0$ results from the normal parts of the Hamiltonian $H$ while $\hat{\Delta}$ originates from the pairing contributions. The Pauli matrices  $\tau_{i}$ (with $i=x,y,z$) operate in  particle-hole space. Diagonalizing the BdG Hamiltonian, the Josephson current can be obtained from $I=( 2 e/\hbar )\partial E_{0}(\phi)/\partial \phi$, where $E_{0}(\phi)$ is the many-body ground state energy. Corresponding results are presented in Fig.\ \ref{plot1}. 

The spectrum of $\hat{\cal H}_{BdG}$ is shown in the lower panels for the topological (left and middle) and 
the nontopological (right) phase. In the topological phase, there are zero-energy bound states (light-blue curves) which  emerge from the Majoranas states localized at the far ends of the finite-length chains. 
The solid red curves emerge from the hybridization of the dot states with the adjacent Majoranas. 
In the nontopological phase, the subgap states are gapped. As a consequence of Kramers theorem, the subgap states are twofold degenerate at $\phi$ equal to integer multiples of $\pi$. At other flux values, time reversal is broken by the phase bias and the subgap states are nondegenerate. The top panel of  Fig.\ \ref{plot1} shows the Josephson current for values of $\mu$ both in the topological 
($\mu<2\lambda$) and the nontopological ($\mu>2\lambda$) phases. In the topological phase, the Josephson current jumps at $\phi=\pi$ (up to finite-size effects), reflecting the level crossing of the subgap states. (Note that we assume complete equilibration over fermion parities.) The nontopological phase exhibits the usual smooth behavior. 
 
  \begin{figure}[t]\begin{center}
  \includegraphics[width=0.9\columnwidth]{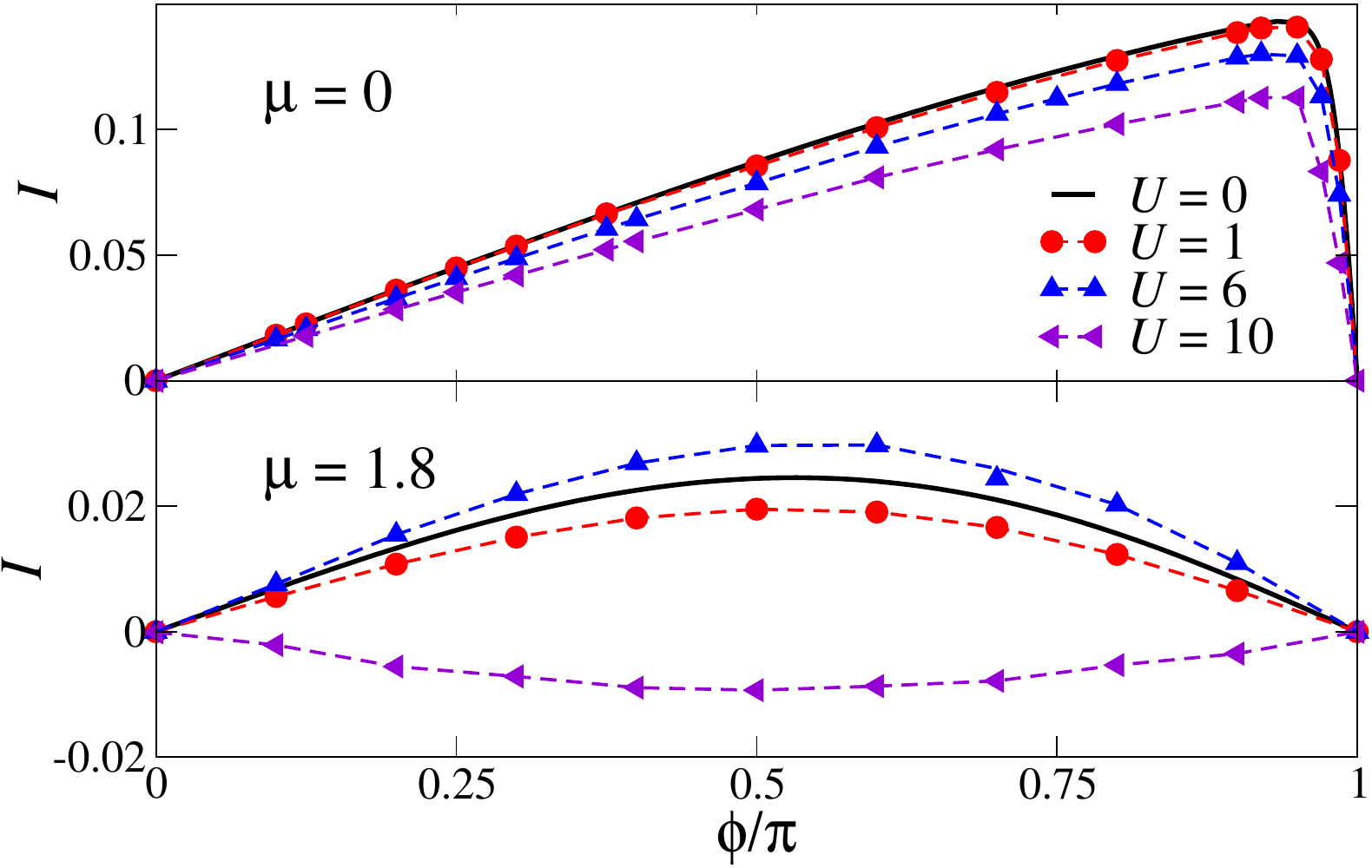}
  \caption{(Color online) Josephson current for an interacting quantum dot at $\beta=400$ with 
  $ t^{\prime}= t$, $\varepsilon_d=-U/2$,
 $\lambda=t/2$, $\Delta=t/5$. The  upper (lower) panel corresponds to  the topological  (nontopological) phase.
 The  values of $U$ and $\mu$ are indicated in the Fig. Energies are expressed in units of $t=1$ .} \label{fig3}
\end{center}\end{figure}
  
For a nonzero interaction $U$, the Josephson current can be calculated by evaluating $G^{(21)}_{d,\sigma}( i \omega_n)$ using quantum Monte Carlo simulations \cite{qmc}. Previous works on S-QD-S junctions 
\cite{fasepi2,fasepi4,fasepi5} as well as normal wires coupled to correlated dots and molecules \cite{lilichac,nos} proved this strategy to be accurate and reliable. We perform a Shiba transformation, mapping $H$ to a particle-number conserving Hamiltonian with negative $U$ \cite{fasepi4}. The Green function of the transformed problem is then calculated by the algorithm introduced in Refs.\ \cite{cont1,cont2}. Inversion of the Shiba transformation leads to ${G}_{d, \sigma}(i \omega_n)$ which enters the Josephson current (\ref{jcur}). Results for a half-filled configuration (i.e., $\langle n_{d \uparrow}+n_{d \downarrow} \rangle =1 $) are shown in Fig.\ \ref{fig3}. 

The nontopological case (bottom panel) shows the expected $0$ to $\pi$ transition. When coupling the quantum dot to superconducting leads, the local moment persists when $\Delta$ is larger than the Kondo temperature $T_K$, but becomes Kondo screened by the quasiparticle states for $\Delta\ll T_K$. For a particle-hole symmetric configuration, the Kondo temperature of the junction is given by $k_B T_K = \sqrt{\delta U/2} \exp{\left( -\pi U/8 \delta \right)}$  \cite{kondo} with the hybridization parameter $\delta \sim\pi  (t^\prime)^2 \sqrt{1-\mu^2/(2t)^2}/(2t)$. Consequently, there is a $0$-$\pi$ transition as $U$ increases. For $U=t$, the dot is in the intermediate valence regime, while for $U=6t$ and $U=10t$, it would be in the Kondo regime when attached to normal leads. In our case, $k_B T_K \sim \Delta$ for $U\sim 8.5 t$, consistent with the observed transition between $0$- and $\pi$-junction behavior between $U=6t$ and $U=10t$. 

It is our central observation that there is no corresponding $0$-$\pi$ transition when the superconducting leads are in the topological regime. Instead, the current-phase relation remains similar to the noninteracting case for all interaction strengths $U$. In particular, the abrupt dependence at a phase difference of $\phi=\pi$, while slightly smoothed by finite temperature, becomes more pronounced as the number of sites increases, as in the noninteracting case (cp.\ inset of Fig.\ 1). These results suggest that the impurity spin is efficiently screened in the topological case, despite the presence of the superconducting gap. This robust screening of the spin of the quantum dot originates from its interaction with the subgap states emerging from the Kramers pairs of Majoranas of the adjacent left and right wires. 
   
{\em Effective Hamiltonian.---}To arrive at this conclusion, we interpret our numerical results in the context of an effective Hamiltonian. Consider a singly-occupied, interacting quantum dot coupled to two time-reversal-invariant topological superconductors. For simplicity, we assume that the superconducting gap is large compared to the Kondo temperature so that we can neglect hybridization with the quasiparticle continuum. Then, we only need to consider the hybridization with the subgap states originating from the Majorana bound states. We can project out the empty and doubly occupied dot states by employing a Schrieffer-Wolff transformation \cite{s-w} (see also \cite{tri13} for similar considerations). This yields an effective Hamiltonian in the eight-dimensional subspace spanned by the two eigenstates of the quantum dot spin ${\bf S}_d$ and the two states for each of the superconducting leads which are associated with the Kramers pair of Majorana operators. Here, we sketch the derivation of this Hamiltonian (for details, see \cite{suppl}, Sec.\ 3).

 \begin{figure}[t]\begin{center}
  \includegraphics[width=\columnwidth]{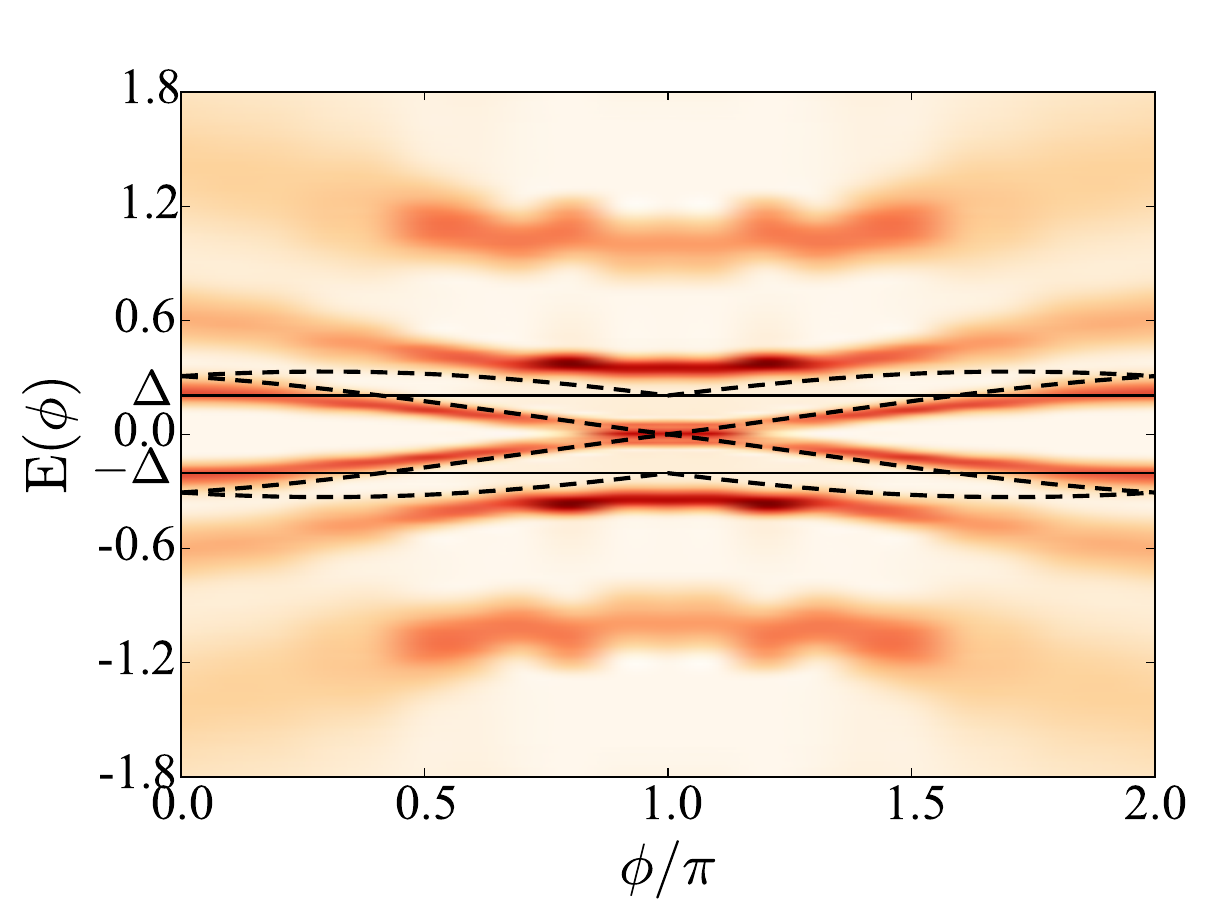}
  \caption{(Color online) LDOS at the quantum dot  obtained by QMC. The black dashed lines are the predictions for the peaks in the density of states on the basis of  $H_{\rm eff}$ with $J=0.2$. The amplitude of the superconducting gap $\Delta=0.2$ is indicated in thin lines. Other parameters are $U=4t$, $J=0.2$, $\lambda=0.5 t$, $t^{\prime}=t$, $\mu=0$ and $\beta=400$. }\label{figenef}
\end{center}\end{figure}

In a first step, we project the tunneling Hamiltonian $H_c$ to the subgap  states  of the  wires, giving 
$H_{c} = 
 t_{\rm eff} e^{i \phi/4}\sum_{\sigma} \left( \Gamma_{L,\sigma}^{\dagger} d_{\sigma}+ d^{\dagger}_{\sigma}  
 \Gamma_{R,\sigma} \right) + {\rm H.c} $, where $t_{\rm eff} \lesssim t^{\prime}$ and 
the Bogoliubov operators for the zero-energy modes  satisfy  $\Gamma_L^{\dagger}=\Gamma_{L,\uparrow}^{\dagger}= 
i \Gamma_{L,\downarrow}$, and $\Gamma_R^{\dagger}=\Gamma_{R,\uparrow}^{\dagger}=-i \Gamma_{R, \downarrow}$ 
(see \cite{footnote} and \cite{suppl}, Sec.\ 2).
 Focusing on the particle-hole symmetric point $\varepsilon_d = -U/2$, and eliminating the empty and doubly occupied states of the quantum dot by a Schrieffer-Wolff transformation, we obtain 
(see \cite{suppl}, Sec.\ 3 for details, including more general configurations)
\begin{eqnarray}
& & H_{\rm eff}= J \left\{ S_d^z \left[ (n_L + n_R -1) + i \sin \frac{\phi}{2} \left( \Gamma_L^{\dagger} \Gamma_R - \Gamma_R^{\dagger} \Gamma_L \right)  \right] \right.\nonumber \\
& & \,\,\,\,\,\,\,\,\left.+ i \cos\frac{\phi}{2} \left( S_d^- \Gamma_L^{\dagger} \Gamma_R^{\dagger} -  S_d^+  \Gamma_R \Gamma_L \right) \right\},
\label{low}
\end{eqnarray}
where $J= 4 t_{\rm eff}^2/ U$ and we defined the occupations $n_\alpha = \Gamma_\alpha^\dagger\Gamma_\alpha$. A convenient basis for this Hamiltonian is $|\sigma, n_L, n_R\rangle$ with $n_\alpha = 0,1$ and $\sigma = \uparrow, \downarrow$. Note that $n_\alpha$ also labels the polarization of the fractional spins. 

The Hamiltonian $H_{\rm eff}$ is easily diagonalized. It conserves the number parity of $n = n_L+n_R$. For $n=1$,  the terms involving $S_d^{\pm}$ do not contribute and we find doubly-degenerate eigenstates which are linear superpositions of $|\sigma,1,0\rangle$ and $|\sigma,0,1\rangle$ with energy $ \pm  J/2  \sin(\phi/2)$. For even occupations $n$, we have two phase-independent states with degenerate eigenergies $J/2 $ corresponding to $|\uparrow,1,1\rangle$ and $|\downarrow,0,0\rangle$ as well as a pair of nondegenerate states with energies $-J/2  \pm J \cos(\phi/2)$, which are linear combinations of the states $|\uparrow,0,0\rangle$ and $|\downarrow,1,1\rangle$. 

At all phase differences, the ground state is an equal-probability superposition of  $|\uparrow,0,0\rangle$ and $|\downarrow,1,1\rangle$. These states describe configurations with overall zero spin. Indeed, in both states the quantum dot spin of $\hbar/2$ is pointing opposite to the fractional spins of $\hbar/4$ of the two adjacent superconductors. Thus, these configurations can be interpreted as an effective singlet configuration of the quantum dot spin and the fractional spins of the topological superconductors. 

Similar to the singlet formation via hybridization with the quasiparticle continuum of nontopological superconductors \cite{kirs}, this singlet formation with the fractional spins quenches the $\pi$-junction behavior. Indeed, the low-energy spectrum emerging from the Schrieffer-Wolff treatment predicts a Josephson energy which is minimal at phase differences equal to integer multiples of $2\pi$. Moreover, we also see that there is a cusp in the ground state energy at a phase difference of $\pi$. Both of these results are consistent with our numerical results which incorporate the hybridization with the quasiparticle continuum above the superconducting gap. 

In Fig.\ \ref{figenef}, we benchmark our low-energy Hamiltonian with results for the local density of states (LDOS) at the quantum dot $\rho(\omega)= -2 \sum_{\sigma} \mbox{Im}[G^R_{d,\sigma}(\omega)]$. 
The latter was calculated by analytically continuing the Monte Carlo data to the real frequency axis. Results are shown in the color plot. The low-energy spectrum obtained from $H_{\rm eff}$ is shown as solid lines for comparison. The peaks in the LDOS reflect the energy necessary to add or remove one particle. Thus, the peak positions can be estimated from $H_{\rm eff}$ by the energy difference between the odd-parity eigenstates and the ground state, which yields
$\pm \left[   J/2 + J |\cos(\phi/2)| \pm J/2 \sin(\phi/2)\right]$. 
We find that our numerics is qualitatively consistent with the predictions of $H_{\rm eff}$, although the numerics is performed in a regime where the addition spectrum already hybridizes with the quasiparticle continuum. Apart from shifts in energy, the hybridization lifts the degeneracies at $\phi=0$ and $2\pi$. Besides the low energy features which are qualitatively described by $H_{\rm eff}$, the numerical results also exhibit high-energy features at $\pm U/2$, which are associated with the charge-transfer peaks of the impurity Anderson model. 

TRITOPS-QD-TRITOPS Josephson junctions combine topological superconductivity with time reversal symmetry and electron-electron interactions. While this is superficially similar to quantum spin Hall Josephson junctions including interactions either within the edge states \cite{zhangkane,schmidt} or through coupling to an interacting quantum dot \cite{peng,sau}, these two types of Josephson junctions are governed by  remarkably different physics. Quantum spin Hall Josephson junctions exhibit an $8\pi$-perodic Josephson effect which can be interpreted as resulting from the tunneling of $e/2$ charges enabled by the formation of $Z_4$ parafermions or from a spin transmutation as a consequence of the fermion parity anomaly \cite{peng}. In contrast, the present system has a Josephson effect which is $4\pi$ periodic and results from an effective singlet formation with two fractional spins.
   
{\em Acknowledgements}. We acknowledge support from CONICET, and UBACyT, Argentina as well as Deutsche Forschungsgemeinschaft and Alexander von Humboldt Foundation, Germany. LA thanks the ICTP-Trieste for hospitality through a Simons associateship. This work was sponsored by PIP 112-201101-00832 of CONICET and PICT 2013-1045, PICT 2012- of the ANPCyT.

\begin{widetext}
 \section{1. Josephson current and Green functions} \label{apa}

Starting from the definition, the Josephson current
\begin{eqnarray} \label{jcur}
I &=& -2 t^{\prime} \sum_{\sigma} \mbox{Im}\left[ \langle c^{\dagger}_{\alpha, 1, \sigma} d_{\sigma}  \rangle \right] ,
\end{eqnarray}
it may be written as
\begin{equation} \label{curg}
I=  - \frac{2 t^{\prime}}{\beta} \sum_{\sigma} \sum_n \mbox{Im} \left[ G^{(11)}_{1\alpha,d,\sigma}(i \omega_n) \right],
\end{equation}
where $G^{(11)}_{1\alpha,d,\sigma}(i \omega_n)$ is the Matsubara component of the matrix element $(1,1)$ of the Nambu-Gorkov imaginary-time Green function
\begin{equation}\label{green}
\hat{G}_{1\alpha ,d,\sigma}(\tau) = 
    \left( \begin{array}{cc} - \langle T_{\tau} \left[ \hat{c}_{\alpha, 1, \sigma} (\tau)
\hat{d}^{\dagger}_{\sigma} (0) \right] \rangle & - \langle T_{\tau} \left[ \hat{c}^{\dagger}_{\alpha, 1, \sigma} (\tau)
\hat{d}^{\dagger}_{\overline{\sigma}} (0) \right] \rangle \\
- \langle T_{\tau} \left[ \hat{c}_{\alpha, 1, \sigma} (\tau)
\hat{d}_{\overline{\sigma}} (0) \right] \rangle & - \langle T_{\tau} \left[ \hat{c}^{\dagger}_{\alpha, 1, \sigma} (\tau)
\hat{d}_{\sigma} (0) \right] \rangle
\end{array} \right). 
\end{equation}
The latter satisfies the following Dyson equation
\begin{equation} \label{dy}
\hat{G}_{1\alpha ,d,\sigma}(\tau) =  \hat{g}_{1\alpha, \sigma}(\tau) \hat{t}_c  \hat{G}_{d, \sigma}(\tau),
\end{equation}
with $\hat{t}_c  =- t^{\prime} \hat{\bf \tau}_3$, being $\hat{\bf \tau}_3$ the third Pauli matrix
and
\begin{equation}\label{green0}
\hat{g}_{1\alpha , \sigma}(\tau) = 
    \left( \begin{array}{cc} - \langle T_{\tau} \left[ \hat{c}_{\alpha, 1, \sigma} (\tau)
\hat{c}^{\dagger}_{\alpha, 1, \sigma} (0) \right] \rangle_0 & - \langle T_{\tau} \left[ \hat{c}^{\dagger}_{\alpha, 1, \sigma} (\tau)
\hat{c}^{\dagger}_{\alpha, 1, \overline{\sigma}} (0) \right] \rangle_0 \\
- \langle T_{\tau} \left[ \hat{c}_{\alpha, 1, \sigma} (\tau)
\hat{c}_{\alpha, 1, \overline{\sigma}} (0) \right] \rangle_0 & - \langle T_{\tau} \left[ \hat{c}^{\dagger}_{\alpha, 1, \sigma} (\tau)
\hat{c}_{\alpha, 1, \sigma} (0) \right] \rangle_0
\end{array} \right).
\end{equation}
Here $\langle É \rangle_0$ denotes the mean value with respect to the density matrix of the disconnected wire, while $\langle É\rangle$ is the mean value
with respect to the density matrix of the full Hamiltonian. 
Substituting (\ref{dy}) in (\ref{curg}) leads to 
\begin{eqnarray} 
I &=& \frac{2 t^{\prime 2}}{\beta} \sum_{\sigma} \sum_{n} \mbox{Im}\left[g_{1\alpha,\sigma }^{(12)}(i \omega_n) G^{(21)}_{d,\sigma}( i \omega_n) \right] .
\end{eqnarray}

\section{2. Continuum model for the TRITOPS wire}
The Hamiltonian proposed in Ref. \cite{fan-kane-mele}, written  in Eq. (1) of the main text, expressed in $k$-space reads
\begin{equation}
\label{wires}
H= \sum_{k,\sigma} \{ \left[-2 t \cos(k a) + 2 \lambda_{\sigma} \sin(k a)  - \mu \right] c^{\dagger}_{k,\sigma} c_{k,\sigma} + 2 \Delta \cos(k a) c^{\dagger}_{k,\sigma} c^{\dagger}_{-k, -\sigma} + H.c \},
\end{equation} 
where $a$ is the lattice constant.
Without the pairing term, this Hamiltonian defines dispersion relations for $\uparrow$ and $\downarrow$ fermions, which are shifted one another along the $k$-axis as a consequence of the spin-orbit term. Hence, there are in general four Fermi vectors, $k_{\alpha, \sigma}$, where $\alpha=l,r$ denotes left and right movers, respectively. They satisfy $k_{r,\sigma}=-k_{l,\overline{\sigma}}$, with
$\overline{\uparrow} = \downarrow$ and $\overline{\downarrow}=\uparrow$.  
 In addition, the pairing term opens a gap 
$\Delta \cos(k)$, which vanishes at $k=\pm \pi/2$ and has different signs for $|k| < \pi/2$ and for $|k|> \pi/2$.  The low-energy model for this Hamiltonian is derived by writing
\begin{equation} \label{linear}
c_{j,\sigma} \sim \sqrt{a} \left[ e^{ik_{r,\sigma} j a } \psi_{r, \sigma} +  e^{-ik_{l,\sigma} j a } \psi_{l, \sigma} \right],
\end{equation}
 where $\psi_{\alpha,\sigma}(x)$, with $\alpha=L,R$ are fermionic fields for the left and right movers, respectively, with spin $\sigma$.
 
Let us focus on the case where $k_{r,\uparrow}=-k_{l,\downarrow}=\pi/2$, which corresponds to the critical case where the superconducting gap vanishes. Instead, the other fermions at the Fermi level with $k_{r,\downarrow}=-k_{l,\uparrow}$ are under the effect of a finite superconducting pairing,  and a gap opens in the spectrum at these points. 
For the derivation of the low-energy model in this particular case we project on the gapless states and (\ref{linear}) reads
 \begin{equation} \label{chiral}
c_{j,\uparrow}  \sim  \sqrt{a} e^{i \pi ja /2  } \psi_{r, \uparrow} , \;\;\;\;\;\;\;\; \;\;\; c_{j,\downarrow}  \sim  \sqrt{a}   e^{-i \pi j a/2  } \psi_{l, \downarrow},
\end{equation}
where the indices $l,r$ are now redundant and will be omitted. 
 Substituting in (\ref{wires}) and, keeping the leading orders for each of the terms of the Hamiltonian, we get the following low-energy continuum model
 \begin{equation} \label{low}
 H \sim \left( 2 \lambda -\mu \right) \sum_{\sigma} \int dx \psi^{\dagger}_{\sigma}(x)   \psi_{\sigma}(x) 
  + 2 t a \sum_{\sigma}\int dx \psi^{\dagger}_{\sigma}(x)  i \partial_ x  s_{\sigma} \psi _{\sigma}(x) 
 + 
 2 \Delta a \int dx \psi^{\dagger}_{\uparrow}(x)  i \partial_ x \psi^{\dagger}_{\downarrow}(x) + {\rm H.c.},
 \end{equation}
 with $s_{\uparrow, \downarrow} = \uparrow, \downarrow$. For simplicity, and without affecting the key features of the original model, 
 we will drop the second term $\propto t$. 
 
 The same arguments above can be repeated for
$k_{r,\downarrow}=-k_{l,\uparrow}=\pi/2$, in which case the pairing between states with $k_{r,\uparrow}=-k_{l,\downarrow}$ have different sign with respect to the ones in previous case. Hence,
we get the Hamiltonian (\ref{low}) but with a $-$ sign in front of the last term. 
The corresponding Bogoliubov-de Gennes matrix for the full low-energy Hamiltonian reads
 \begin{equation} \label{bdg}
 {\cal H}_{TRITOPS} = (2 \lambda - \mu) \tau_z - 2 a \Delta p \tau_x \sigma_z,
 \end{equation}
 where $\tau_{x,y,z}$ are Pauli matrices acting on the spinors $\psi_{p \uparrow}^{\dagger}= \left(c^{\dagger}_{p \uparrow} \;\;c_{-p \downarrow}  \right)$, while $\sigma_z$ acts on the spin degrees of freedom. The topological phase takes place for 
 $2 \lambda - \mu >0$, while $ 2 \lambda - \mu <0$, corresponds to a trivial superconductor. If we define a domain wall where $2 \lambda - \mu$ changes sign, we get zero-energy modes 
 localized at the domain wall. 
 To get the structure of the zero-energy modes, we consider a linear domain wall centered at $x=0$, i.e.,  $2 \lambda(x)-\mu(x)=\alpha x$. For $\alpha >0$, we have a nontopological
 (topological) superconducting phase for $x<0$ ($x>0$), while the opposite situation takes place for $\alpha<0$. 
 Following Ref. \cite{felix} we can calculate the zero-energy modes localized at the wall by calculating $H^2$ and identifying it with the Hamiltonian for an Harmonic oscillator. 
 The solutions for the zero-modes read
 \begin{equation}
 \Gamma_{\uparrow}  =   \int dx \phi_0(x) \left[ \psi_{\uparrow}(x) \pm i \psi^{\dagger}_{\downarrow}(x) \right], \;\;\;\;\;\;\;\;\;\;\; 
        \Gamma_{\downarrow}  =  \int dx \phi_0(x) \left[ \psi_{\downarrow}(x) \pm i \psi^{\dagger}_{\uparrow}(x) \right], 
 \end{equation}
 where the $\pm$ corresponds to $\alpha <,> 0$, respectively and $\phi_0(x)$ is the zero-mode wave function of the effective harmonic-oscillator Hamiltonian. If we now represent the junction between the TRITOPS wires with a 
 domain wall  at the left with $\alpha >0$ followed by another at the right 
 with $\alpha <0$. The corresponding zero-modes will have the structure of the Bogoliubov operators 
  \beq\label{prop}
 \Gamma_L^{\dagger}=\Gamma_{L,\uparrow}^{\dagger}= i \Gamma_{L,\downarrow},\;\;\;\;\;\; \Gamma_R^{\dagger}=\Gamma_{R,\uparrow}^{\dagger}=-i \Gamma_{R, \downarrow}.
\edq

\section{3. Schrieffer-Wolff transformation and derivation of the effective
low-energy Hamiltonian}

In this Section, we derive the effective Hamiltonian $H_{\mathrm{eff}}$ in
the limit of occupation 1 at the dot for arbitrary coupling to the left and
right leads $t_{L},t_{R}\ll -\varepsilon _{d},U+\varepsilon _{d}$. This is
analogous to the derivation of a Kondo Hamiltonian from the Anderson model 
\cite{kondo,s-w}

Projecting the operators of the left and right superconductors attached to
the quantum dot onto the low-energy subgap modes leads to the following
Hamiltonian to describe the low-energy physics 
\beq
\label{low2} 
H_{\textrm{low}}= -e^{i\phi /4}\sum_{\sigma } \left( t_{L}\Gamma _{L,\sigma }^{\dagger
}d_{\sigma }+t_{R}d_{\sigma }^{\dagger }\Gamma _{R,\sigma }\right)  + \textrm{%
H.c.} +H_{d}, 
\edq 
where $t_{L},t_{R}\lesssim t^{\prime }$ represent the
coupling to the TRITOPS leads and $\Gamma _{\alpha ,\sigma }$ are the
Bogoliubov operators corresponding to the zero-modes of the wires localized
at the left and right sides of the quantum dot which satisfy the properties (\ref{prop}). 

In the limit of $t_{L},t_{R}\ll -\varepsilon _{d},U+\varepsilon _{d}$, 
we can eliminate the high-energy states
of the quantum dot by recourse to a Schrieffer-Wolff transformation \cite{s-w}.
Since $\Gamma _{\alpha ,\uparrow }$ and $\Gamma _{\alpha ,\downarrow }$ are
related by Eqs. (\ref{prop}), we first write the Hamiltonian $H_{\textrm{low}}$
in terms of the independent operators $\Gamma _{L,\uparrow }=\Gamma _{\uparrow }$
and $\Gamma _{R,\downarrow }=\Gamma _{\downarrow }$. This choice leads to a form
of $H_{\mathrm{eff}}$ with spin-spin interactions similar to the usual Kondo
model. 

Following the conventional procedure  \cite{kondo} we obtain 
\begin{eqnarray} \label{hef}
H_{\mathrm{eff}} &=&-(J_{LL}n_{d \uparrow }+J_{RR}n_{d \downarrow
})/2+S_{d}^{z}[J_{LL}n_{\uparrow }-J_{RR}n_{\downarrow }+J_{LR}\sin (\phi
/2)\left( \Gamma _{\uparrow }^{\dagger }\Gamma _{\downarrow }^{\dagger }+\Gamma
_{\downarrow }\Gamma _{\uparrow }\right) ]  \nonumber \\
&&+J_{LR}\cos (\phi /2)\left( S_{d}^{+}S_{\Gamma }^{-}+S_{d}^{-}S_{\Gamma
}^{+}\right) -iW\cos (\phi /2)\left( \Gamma _{\uparrow }^{\dagger }\Gamma
_{\downarrow }^{\dagger }-\Gamma _{\downarrow }\Gamma _{\uparrow }\right) .
\label{heff2}
\end{eqnarray}
with 
\beq 
J_{\alpha \beta }= \frac{2Ut_{\alpha }t_{\beta }}{(\varepsilon
_{d}+U)(-\varepsilon _{d})} , \ \ \ \ \ \  W= \frac{(2\varepsilon
_{d}+U)t_{L}t_{R}}{(\varepsilon _{d}+U)(-\varepsilon _{d})} .
\edq 
The spin operators are $S_{\Gamma }^{z}=\left( n_{\uparrow }-n_{\downarrow
}\right) /2$, with $n_{\sigma }=\Gamma _{\sigma }^{\dagger }\Gamma _{\sigma }$,~ $%
S_{\Gamma }^{+}=\Gamma _{\uparrow }^{\dagger }\Gamma _{\downarrow }$, ~$S_{\Gamma
}^{-}=\Gamma _{\downarrow }^{\dagger }\Gamma _{\uparrow }$ and similar expressions
for $S_{d}^{z}$ and $S_{d}^{+,-}$. This Hamiltonian conserves the parity
number and total spin projection  $S^{z}=S_{d}^{z}+S_{\Gamma }^{z}$. Eq. (7) in the main text corresponds to (\ref{hef}) for $t_{LL}=t_{RR}=t$ and $\varepsilon_d=-U/2$.

Restricting for the moment to the subspace with total even number of
particles [for which the terms of Eq. (\ref{heff2}) containing $\Gamma
_{\uparrow }^{\dagger }\Gamma _{\downarrow }^{\dagger }$ and $\Gamma _{\downarrow
}\Gamma _{\uparrow }$ become irrelevant], in the case of symmetric coupling
where $J_{LL}=J_{RR}=J_{LR}=J$, the Hamiltonian 
reduces to the Heisenberg Hamiltonian with antiferromagnetic coupling $2J$
along $z$ and $J\cos (\phi /2)$ on the $x-y$ plane. Interestingly, the
exchange interaction takes place between the $1/2$-spin localized at the
quantum dot and an effective $1/2$-spin formed by the $1/4$-spin excitations
localized at the ends of the chains close to the quantum dot. In the general
case, the  lowest-energy eigenstates and can be written as $|\psi _{e}^{\pm
}\rangle =\left( |\uparrow \downarrow \rangle \pm |\downarrow \uparrow
\rangle \right) /\sqrt{2}$, where the first entry corresponds to the spin of
the dot and the second one to the effective spin of the side-chains. The
corresponding eigenenergies are: 
\beq \label{ev}
E_{e}^{\pm } = -(J_{LL}+ J_{RR})/2 \pm
J_{LR}\cos (\phi/2). 
\edq 
The two additional eigenstates are $|\sigma \sigma \rangle $,
with $\sigma= \uparrow, \downarrow$ and have zero energy. 

In the subspace with odd number of particles, the eigenstates are a mixture
of $|\sigma ,0\rangle $ and $|\sigma ,\uparrow \downarrow \rangle $, with
eigenenergies 
\beq \label{od}
E_{o}^{\pm } = - \frac{J_{LL}+J_{RR}}{4} \pm \sqrt{\left[ 
\frac{J_{LL}-J_{RR}}{4}\right] ^{2}+\frac{J_{LR}^{2}}{4}\sin ^{2}(\phi
/2)+W^{2}\cos ^{2}(\phi /2)} 
\edq 
and they are doubly degenerate due to
the two possible orientations of $\sigma $ at the dot.

The eigenenergies given in the main text correspond (up to an energy shift of $-J/2$) to Eqs. (\ref{ev}) and (\ref{od}) for $J_{LL}=J_{RR}=J$ and $W=0$.

For arbitrary $\phi $, the GS of the full Hamiltonian is in the even
subspace. For $\phi =(2m+1)\pi $, with $m$ integer, the states $|\psi
_{e}^{\pm }\rangle $ cross and they become also degenerate with the doublet 
of the odd subspace, hence, defining a $4$-fold degenerate level crossing.
In all the cases, this main features do not depend on the value of $
\varepsilon _{d}$ and $t_{L},t_{R}$. Due to this level crossing, the Josephson current has a peridicity of $4 \pi$.

The diagonalization of the effective Hamiltonian $H_{\mathrm{eff}}$ leads to
the 8 lowest-energy states of the spectrum of $H_{\mathrm{low}}$. The latter
Hamiltonian has 8 additional higher energy states corresponding to the dot
empty or doubly occupied.

The effect of a magnetic field $B$ locally applied at the quantum dot can be analyzed in the limit of $B \ll U$ 
by adding to the effective Hamiltonian of Eq. (\ref{heff2}) 
a term ${\bf B} \cdot {\bf S_d}$. In the case of a magnetic field in the direction of the spin-orbit interaction, ${\bf B} = B {\bf e}_z$, the eigenenergies of Eq. (\ref{ev}) are modified to
\beq \label{ev1}
E_{e}^{\pm } = -\frac{J_{LL}+ J_{RR}}{2} \pm \frac{1}{2}\sqrt{ B^2 + 4 J_{LR}^2 \cos^2 (\phi/2)},
\edq 
while the eigenenergies (\ref{od}) change to 
\beq \label{od1}
E_{o} = - \frac{J_{LL}+J_{RR}}{4} \pm \frac{B}{2} \pm \sqrt{\left(
\frac{ J_{LL}-J_{RR}}{4} \right)^{2}+\frac{J_{LR}^{2}}{4}\sin ^{2}(\phi
/2)+W^{2}\cos ^{2}(\phi /2)}.
\edq 
This implies the lifting of $\uparrow, \downarrow$ degeneracies of the eigenstates and the opening of a gap at $\phi=\pi$. The consequence of the gap opening is a change in the periodicity 
of the Josephson current to $2 \pi$, as in junctions of non-topological superconductors. The ground state remains in the even space and the behavior of the Josephson current corresponds to the
$0$-phase. 

In the case of a magnetic field perpendicular to the direction of the spin-orbit interaction 
${\bf B}=B {\bf e}_x$ the ground state is also in the even subspace. Some degeneracies are lifted but the four-fold degeneracy at $\phi=\pi$ remains and the Josephson current preserves the $4\pi$-degeneracy 
 and the behavior of the Josephson current corresponds to the
$0$-phase.

 \begin{figure}[t]\begin{center}
  \includegraphics[width=0.3\columnwidth,height=4.5cm]{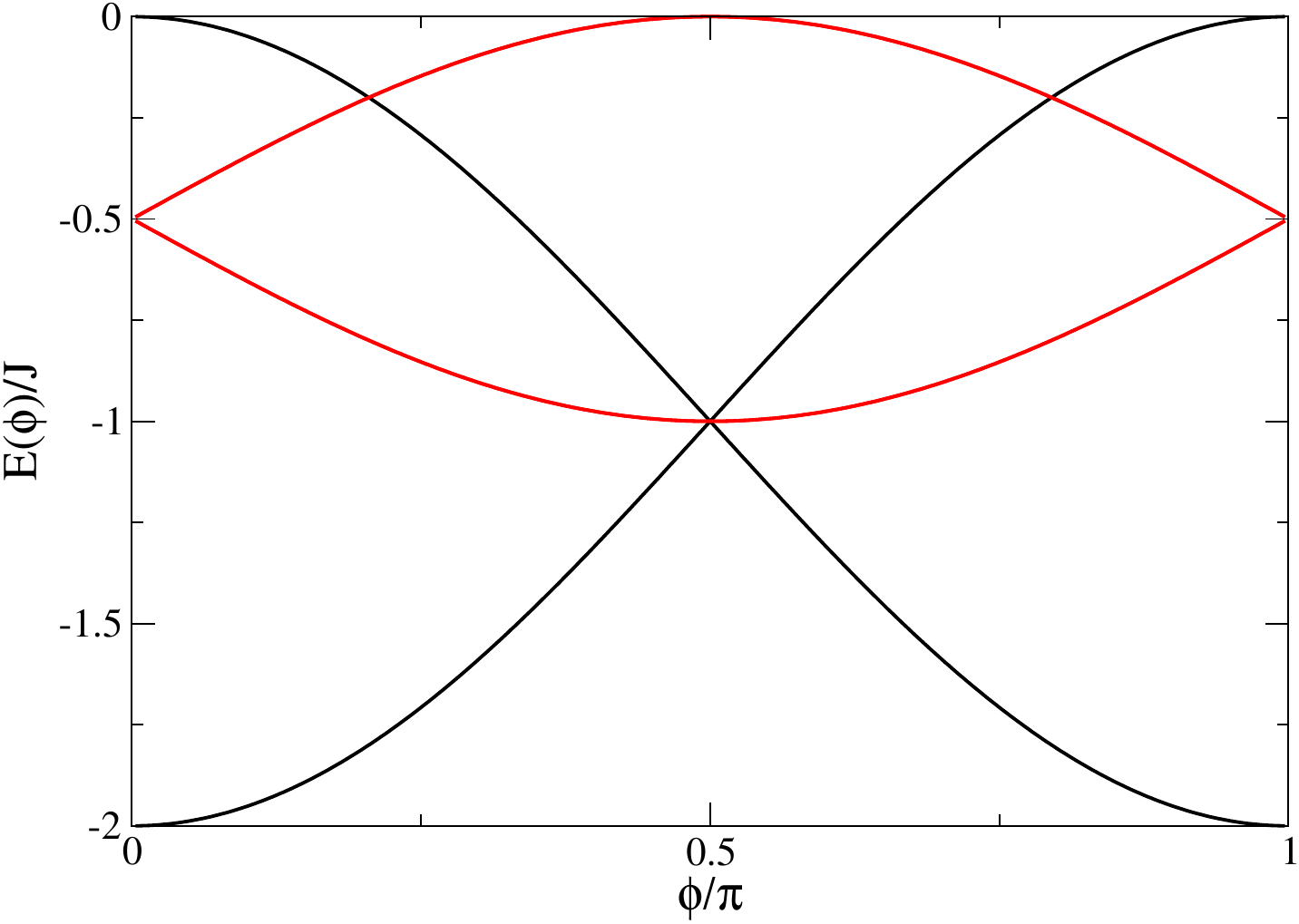},
   \includegraphics[width=0.3\columnwidth,height=4.5cm]{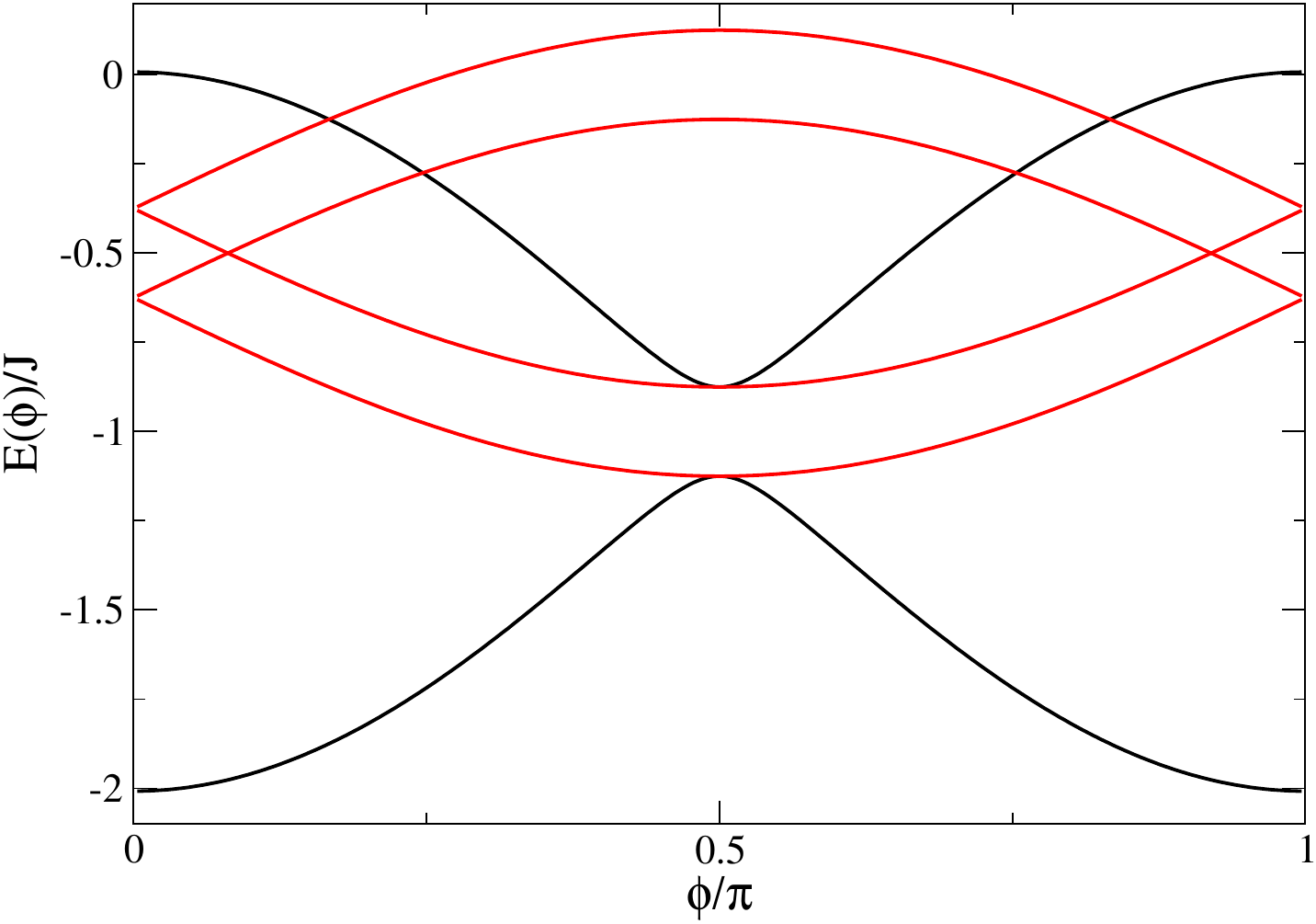},
     \includegraphics[width=0.3\columnwidth,height=4.5cm]{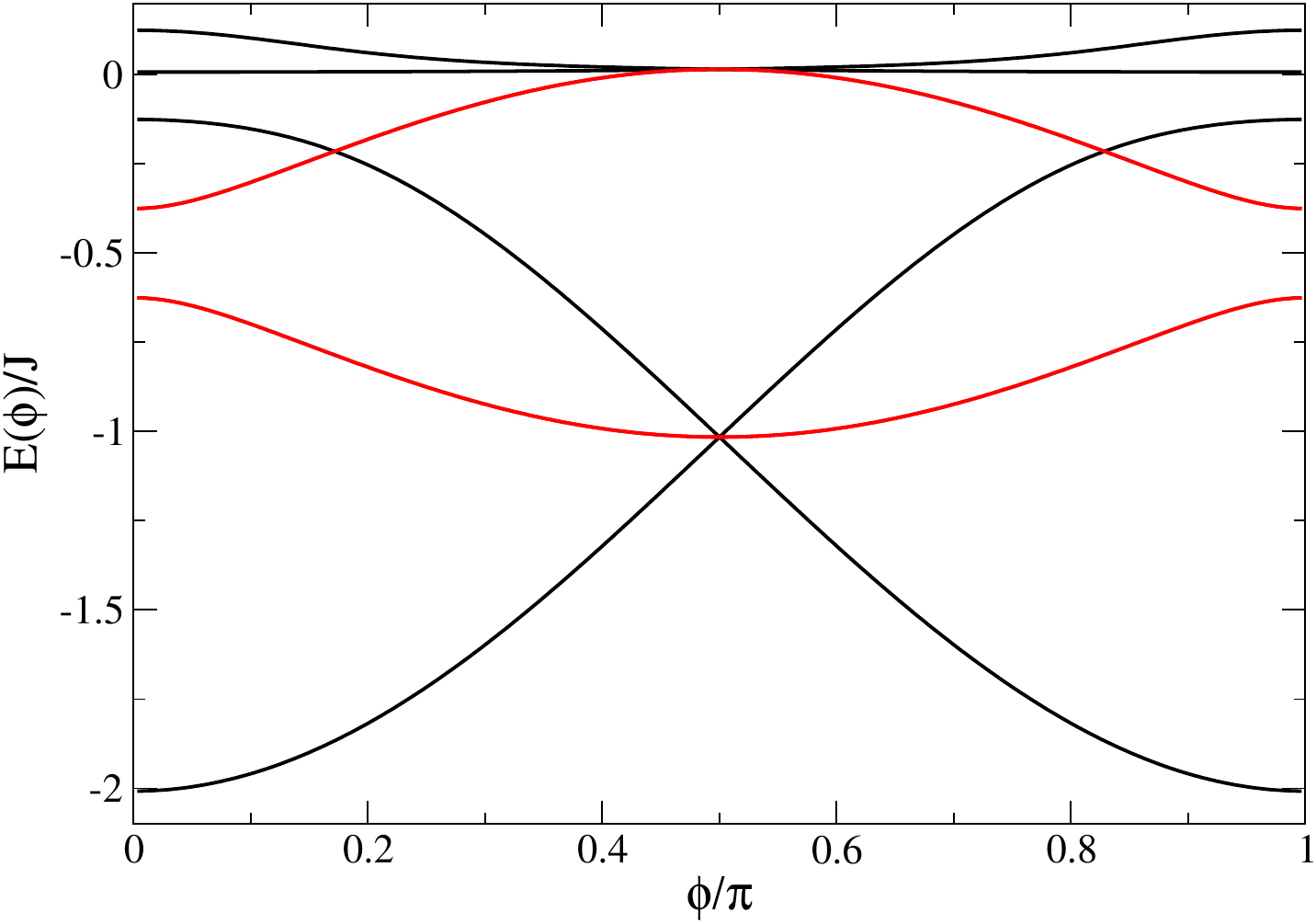}
  \caption{(Color online) Spectrum of $H_{\rm eff}$ for the quantum dot at half filling $\varepsilon_d=-U/2$ and symmetric coupling to the leads 
  $t_{LL}=t_{RR}$, corresponding to $J=1,\; W=0$. Left: ${\bf B}=0$, center ${\bf B}=B {\bf e}_z$ and right ${\bf B}=B {\bf e}_x$. Black and red corresponds, respectively to even and odd subspaces. }\label{plot1}
\end{center}\end{figure}

Results for ${\bf B}= B{\bf e}_z$ and ${\bf B}= B {\bf e}_x$ are shown in Fig. \ref{plot1} for the half-filled configuration and in Fig. \ref{plot2} away from half-filling. The non-symmetric coupling to the leads 
$t_{LL} \neq t_{RR}$ does not change qualitatively these pictures.

\begin{figure}[t]\begin{center}
  \includegraphics[width=0.3\columnwidth,height=4.5cm]{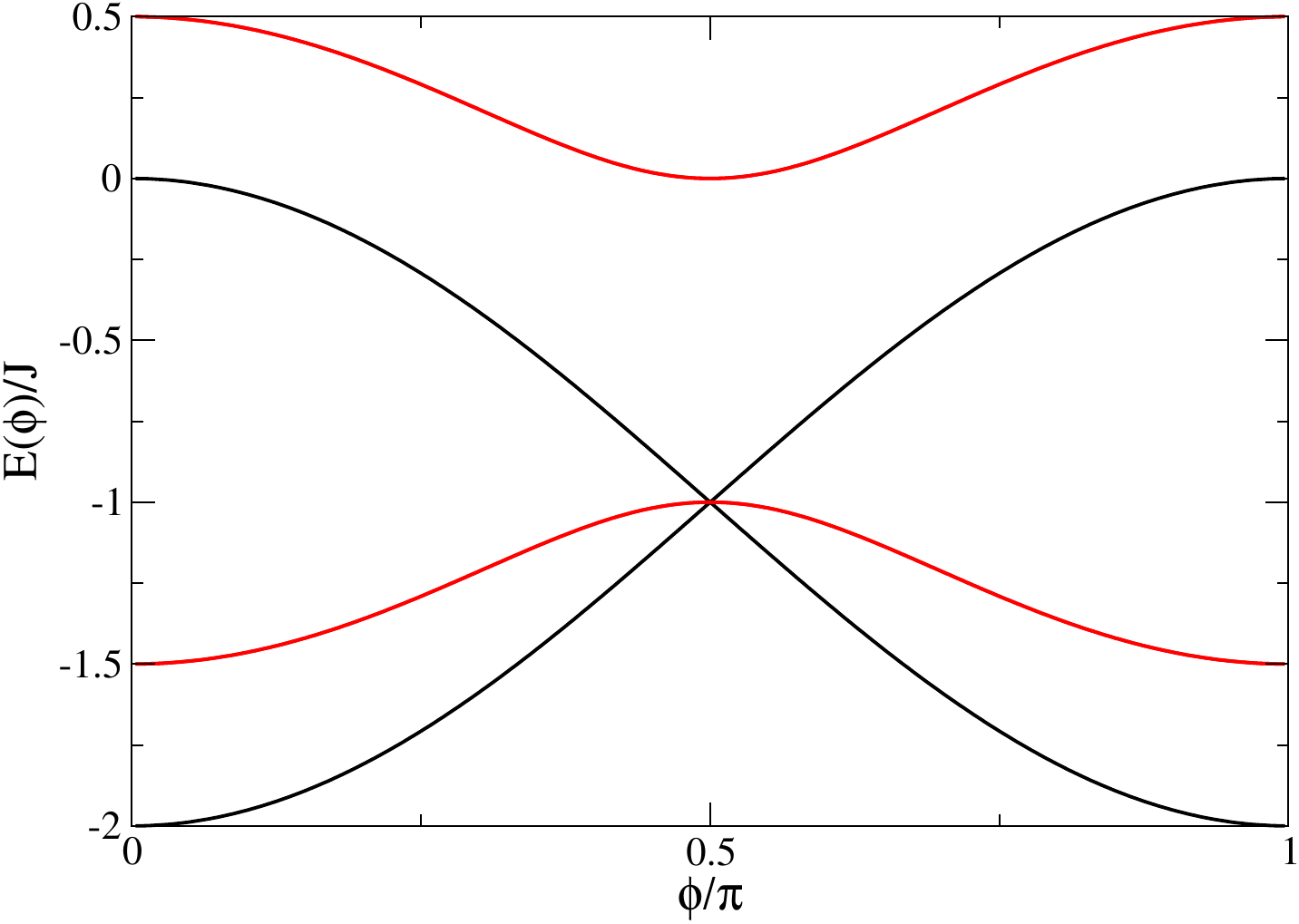},
   \includegraphics[width=0.3\columnwidth,height=4.5cm]{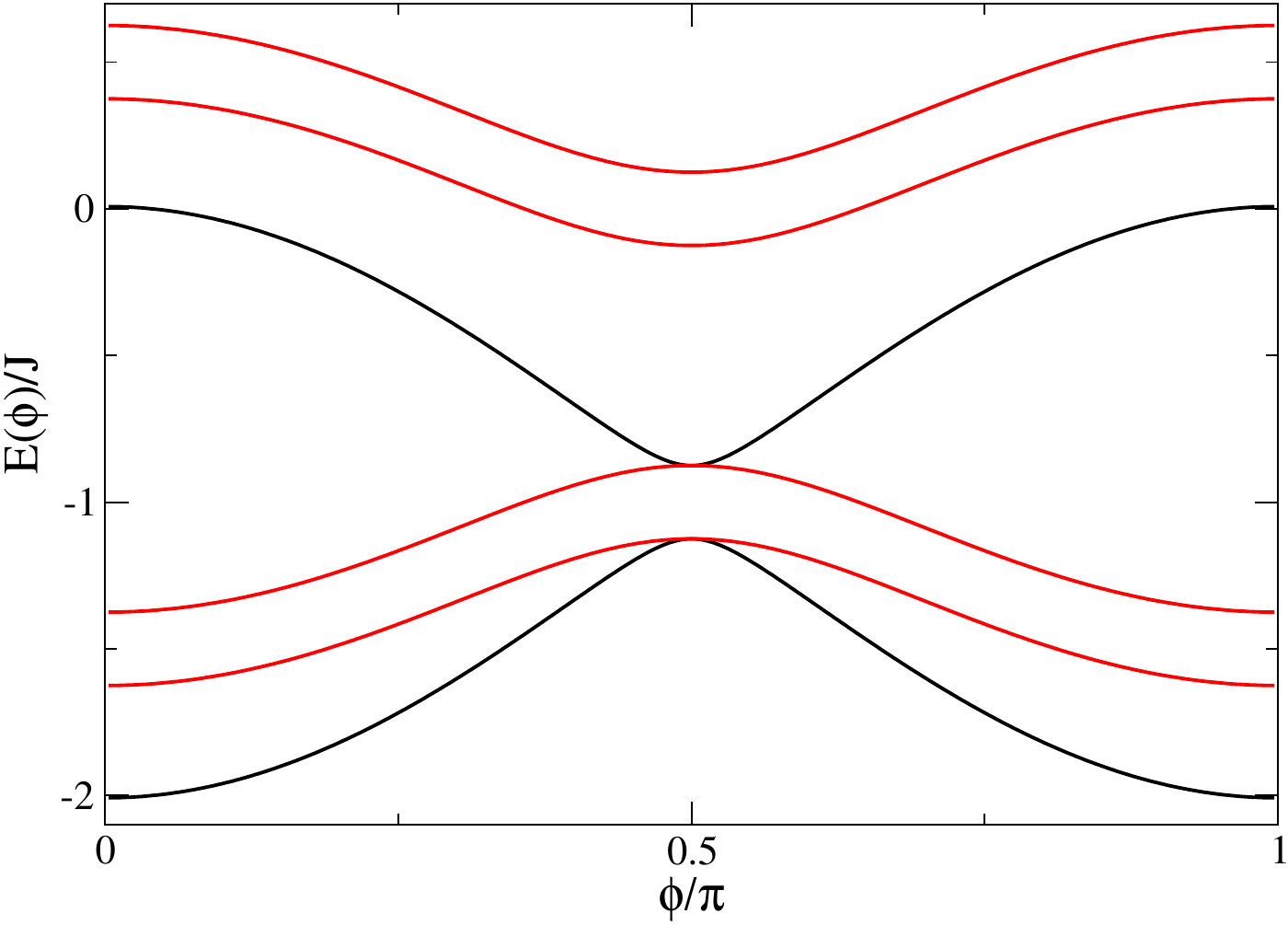},
     \includegraphics[width=0.3\columnwidth,height=4.5cm]{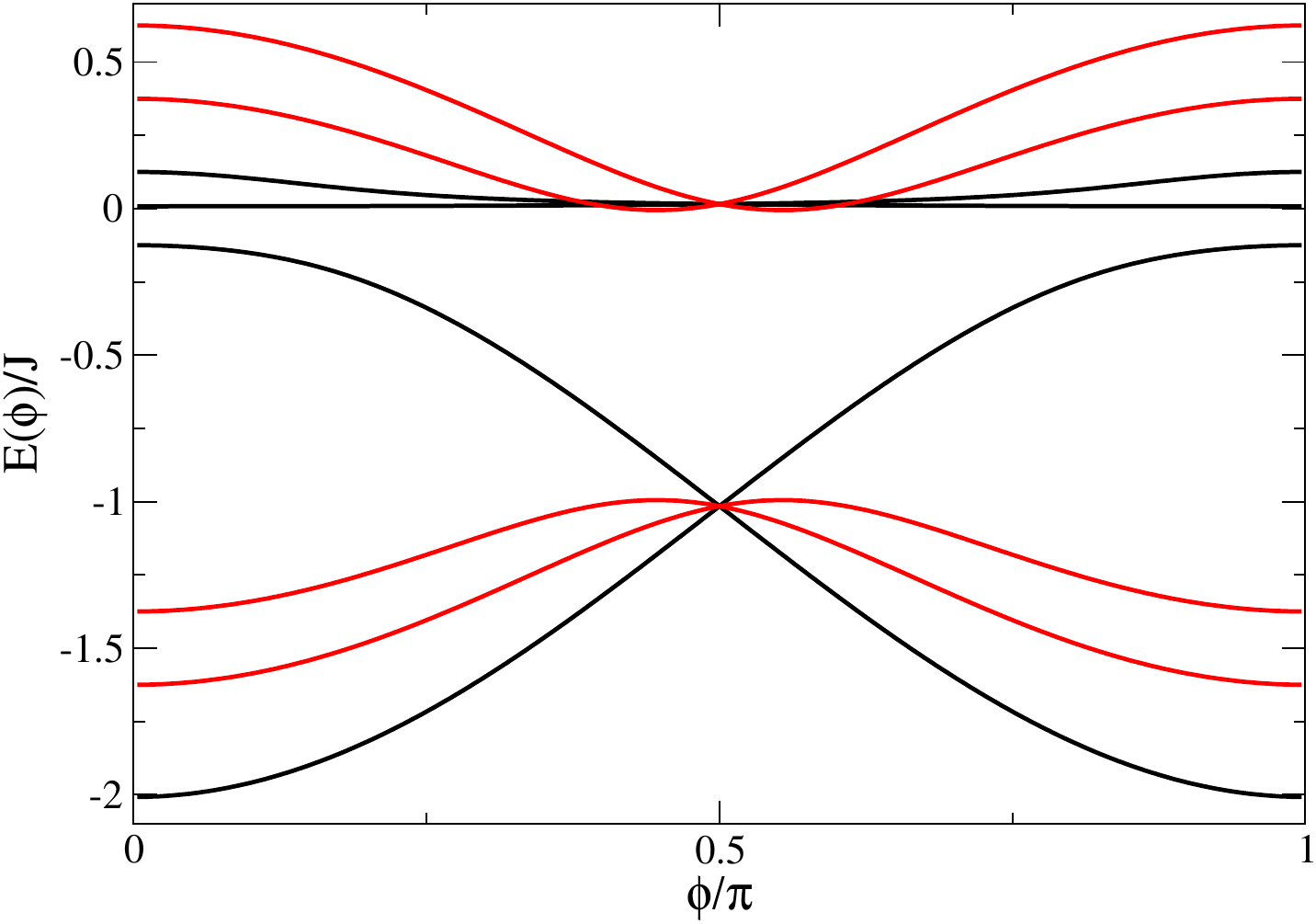}
  \caption{(Color online) Spectrum of $H_{\rm eff}$ for the quantum dot away from half-filling and symmetric coupling to the leads 
  $t_{LL}=t_{RR}$, corresponding to $J=W=1$. Left: ${\bf B}=0$, center ${\bf B}=B {\bf e}_z$ and right ${\bf B}=B {\bf e}_x$. Black and red corresponds, respectively to even and odd subspaces. }\label{plot2}
\end{center}\end{figure}

\end{widetext}


\begin{thebibliography}{99}
\bibitem{footnote} We introduce the notation $\Gamma_{L/R,\sigma}$ to make the tunneling Hamiltonian more compact. It also turns out that this notation is helpful in performing the Schrieffer-Wolff transformation (see \cite{suppl}, Section 3).
\end{thebibliography}
\begin{thebibliography}{99}
\bibitem{note2}We consider a general situation where the coupling to left and right may be different.
\end{thebibliography}
\end{document}